\documentclass[12pt]{article}

\usepackage[a4paper,text={17.8cm,22.4cm}]{geometry}
\usepackage{amsmath,amsfonts,slashed,amssymb,tikz,bm,psfrag,graphicx,xcolor,dsfont}
\usepackage{multicol}
\usepackage[colorlinks=true,linkcolor=red,citecolor=blue]{hyperref}
\usepackage{multirow}
\RequirePackage[sort&compress,square,comma,numbers]{natbib}
\allowdisplaybreaks
\usepackage{multicol}
\usepackage{multirow}
\usepackage{color}

\begin{document}

\begin{titlepage}

\begin{flushright}
%\normalsize
%UWTHPH 2015-30 \\
% arXiv:1505.nnnnn
% v1:
%November 30, 2015
\end{flushright}

\vspace{0.1cm}
\begin{center}
\Large\bf
  $\Lambda_b \to \Lambda_c$ Form Factors from QCD Light-Cone Sum Rules
\end{center}

\vspace{0.5cm}
\begin{center}
{\bf Yan Miao$^a$, Hui Deng$^a$, Ke-Sheng Huang$^b$, Jing Gao$^c$, Yue-Long Shen$^{a}$\footnote{Email: shenylmeteor@ouc.edu.cn, corresponding author} %Fu-Sheng Yu$^b$
} \\
\vspace{0.7cm}
{\it
${}^a$\, College of Physics and Photoelectric Engineering,
Ocean University of China, Qingdao 266100,  China}\\
{\it $^b$School of Nuclear Science and Technology, and Frontiers Science
Center for Rare Isotopes, Lanzhou University, Lanzhou 730000, China}\\
{\it $^c$School of Physics, Nankai University, Weijin Road 94, 300071 Tianjin, China}
\end{center}

\vspace{0.2cm}
\begin{abstract}
In this work, we calculate the transition form factors of $\Lambda_b$ decaying into $\Lambda_c$ within the framework of light-cone sum rules with the distribution amplitudes (DAs) of $\Lambda_b$-baryon. In the hadronic representation of the correlation function, we have isolated both the $\Lambda_c$ and the $\Lambda_c^*$ states so that the  $\Lambda_b \rightarrow \Lambda_c$ form factors can be obtained without ambiguity.  We investigate the P-type and A-type current to interpolate the light baryons for a comparison since the interpolation current for the baryon state is not unique. We also employ three parametrization models for DAs of $\Lambda_b $ in the numerical calculation. We present the numerical predictions on the $\Lambda_b \rightarrow \Lambda_c$ form factors and the branching fractions, the averaged forward-backward asymmetry , the averaged final hadron polarization  and the averaged lepton polarization  of the $\Lambda_b \to \Lambda_c \ell\mu$ decays, as well as the ratio of branching ratios $R_{\Lambda_c}$, and the predicted $R_{\Lambda_c}$ can be consistent with the LHCb data.

\end{abstract}
\vfil

\end{titlepage}

\section{Introduction}

Heavy hadron decays provide an ideal platform to the  precision test of the unitarity of the Cabbibo-Kobayashi-Maskawa(CKM) matrix, to investigate the  CP violation in the standard model(SM) and to search for the new physics signal beyond the SM.  In the past few years,  some unexpected anomalies presented at the observables $R_D,R_{D^\ast}$ of the semileptonic $B$ decays induced by the  $b\to c\ell\nu$ transition, where SM predictions deviate from the data at the $(2 \sim 3)\sigma$ level (see e.g.\cite{HFLAV:2016hnz, BaBar:2012obs, BaBar:2013mob, Belle:2015qfa,Belle:2016ure,Belle:2016dyj, LHCb:2015gmp, LHCb:2017smo,Belle:2019gij,LHCb:2017vlu} ), and the QED corrections also cannot give rise to large corrections to $R_D,R_{D^\ast}$\cite{Beneke:2021jhp}. These anomalies might relate to the violation of the lepton flavour universality (LFU), which is the hint of the existence of new physics(NP) signal.  Therefore, many NP models  have been proposed to explain such tensions, such as $W'$ models, leptoquark models, and models with charged Higgs, see \cite{Bifani:2018zmi,Li:2018lxi} and the references therein. Except for the $B\to D(D^\ast)\ell\nu$ decays, the weak decays of heavy baryon such as $\Lambda_b\to \Lambda_c\ell \nu$ are also mediated by the $b\to cl\nu$ transition, which may provide more hints to make this ``anomaly" more transparent, for a review, see\cite{Bernlochner:2021vlv}.

The fundamental ingredients in the semileptonic  $\Lambda_b\to \Lambda_c\ell \nu$ decays are the $\Lambda_b\to \Lambda_c$ transition form factors. For the heavy-to-heavy transition processes, the heavy quark effective theory(HQET)
\cite{Eichten:1989zv,Georgi:1990um,Manohar:2000dt}provides a natural theoretical framework to analyze the form factor relations and to estimate the power corrections. In the heavy quark limit, only one single independent form factor, namely Isgur-Wise function $\xi(v\cdot v')$\cite{Isgur:1989vq}, appears in
the $\Lambda_b\to \Lambda_c$ transition, and $\xi(v\cdot v')$ satisfies the normalization condition $\xi(1)=1$. The heavy quark symmetry works well in the small recoil region where the $\Lambda_c$ baryon is almost static, and the Lattice QCD simulation based on the first principle is also very suitable to be applied in this region. The predictions of the $\Lambda_b \to \Lambda_c$ form factors with Lattice QCD are given in \cite{Detmold:2015aaa}, and one has to employ phenomenological models to extrapolate the result to the whole momentum region.
To reduce the model dependence, it is very meaningful to calculate the form factor at the large recoil region directly. There already exist some studies using various approaches, such as the quark models\cite{Albertus:2004wj,Faustov:2016pal,Zhao:2018zcb,Zhu:2018jet,Ke:2019smy,Becirevic:2020nmb,Thakkar:2020vpv}, the perturbative QCD approach (PQCD)\cite{Shih:1999yh}, and the combination of HQET and PQCD\cite{Guo:2005qa}.

The QCD sum rules method is a popular approach to evaluated the hadronic parameters according to the quark-hadron duality ansatz. The three-point QCD sum rules have been widely used in the study on the transition form factors.  The $\Lambda_b\to \Lambda_c$ and $\Xi_b\to \Xi_c$ transition form factors have been studied  with the three-point QCD sum rules in \cite{Grozin:1992mk,Dai:1996xv,MarquesdeCarvalho:1999bqs,Wang:2003it,Huang:2005mea,Azizi:2018axf,Zhao:2020mod}. For the heavy-to-light form factors, the light-cone sum rules(LCSR) is more appropriate because the light-cone dominance of the correlation functions is proved at the large recoil region. In the $\Lambda_b \to \Lambda_c$ decays, the final state $\Lambda_c$ moves very fast in the large recoil region, thus the light-cone OPE is applicable. In this paper, we will start from the correlation function defined by the matrix element with the time-ordered product of the $b\to c$ transition weak current and the interpolation current of $\Lambda_c$ baryon sandwiched between the vacuum and the $\Lambda_b$ state, as proposed in\cite{Khodjamirian:2005ea,Khodjamirian:2006st,DeFazio:2005dx}. This heavy-hadron LCSR has been employed to study various decays channels of $B$-meson\cite{Wang:2015vgv,Shen:2016hyv,Wang:2017jow,Lu:2018cfc,Gao:2019lta,Shen:2021yhe}, and the decays of $\Lambda_b$ baryon\cite{Wang:2009hra,Feldmann:2011xf,Wang:2015ndk,Huang:2022lfr}. Similar to $B \to D$ decays, the LCSR with $\Lambda_b$-DAs is valid approximately in the momentum region $0\leq q^2\leq 8$GeV$^2$. In \cite{Wang:2009yma},  the Isgur-Wise function in the $\Lambda_b \to \Lambda_c$ transitions have been studied using the LCSR with $\Lambda_b$-DAs, and a recent study on these form factors are presented in \cite{Duan:2022uzm}. In this paper, we will make improvement on the following aspects:
\begin{itemize}
\item Most of the previous studies concentrate on the Isgur-Wise function which arises in the heavy quark limit. For the physical form factors, especially at the large recoil region of  the final state $\Lambda_c$, there exists large power corrections from the expansion on $1/m_c$. In the present work, we do not perform the heavy quark expansion on the charm quark field, and take advantage of the charm quark field in full QCD to construct the intepolation current of $\Lambda_c$ baryon in the correlation function.
\item We will employ the full set of the three-particle DAs of $\Lambda_b$ up to twist-5, which is accomplished in\cite{Bell:2013tfa}, where  the projector of the DAs in the momentum space is also presented. Since the models of the DAs of $\Lambda_b$-baryon is not well established, we will adopt three different models, i.e., the QCDSR model which is constructed based on QCD sum rules, the exponential model and the free parton model which is proposed by mimicking the $B$-meson DAs  for a comparison.
\item In the previous studies, the heavy $b$ quark is expanded in HQET and only leading power contribution is considered. To improve the accuracy of our predictions,  we will include the $1/m_b$ corrections to the heavy quark field in HQET in the present work.
\item When evaluating the correlation function on the hadronic representation, we insert not only the $\Lambda_c$ meson, but also the parity odd counter particle of $\Lambda_c$ baryon, which can help us to extract the form factor without ambiguity by solving the equation of the obtained sum rules.
\end{itemize}

This paper is organized as follows:   In the next section, we calculate the analytic expression of the $\Lambda_b \to \Lambda_c$ form factors with the $\Lambda_b$-LCSR at tree level, and investigate the power suppressed contribution from the power suppressed heavy quark field. In section 3, we will present the numerical results of the form factors and the experimental observations. We summarize this work in the last section.

\section {The LCSR of  $\Lambda_b \to \Lambda_c$ form factors }
\label{section: tree-level LCSR}

%\subsection{ $\Lambda_b \to \Lambda_c$ form factors}
\label{subsection: form factor definition}
The  heavy-to-light  $\Lambda_b \to \Lambda_c$ form factors induced by $V-A$ current are defined as
\begin{align}
&\langle \Lambda_{c}(p',s')|\bar{c}\gamma_{\mu}(1-\gamma_{5})b|\Lambda_{b}(P,s)\rangle\nonumber \\ =&\bar{u}(p',s')\left[f_{1}(q^{2})\gamma_{\mu}+f_{2}(q^{2})v_{\mu}+f_{3}(q^{2}){v'_{\mu}}
\right]u(P,s)\nonumber \\
 -&\bar{u}(p',s')\left[g_{1}(q^{2})\gamma_{\mu}+g_{2}(q^{2})v_{\mu}+g_{3}(q^{2}){v'_{\mu}}
\right]\gamma_{5}u(P,s).\label{eq:parametrization_simple}
\end{align}
where $P, s$ and $p',s'$ are momentum and spin
of the initial state and final state baryon respectively, $v'=p'/m_{\Lambda_c}$, $q=P-p'$ is the momentum transfer. In the heavy quark limit, i.e.,  $m_b,m_c \to \infty$, the form factors $f_1$ and $g_1$ reduce to one unique Isgur-Wise function $\zeta(w)$ , where $w=v\cdot v'$, and $f_2=f_3=g_2=g_3=0$.  At zero recoil limit $v\cdot v'=1$, we have the normalization condition $\zeta(1)=1$. Since in this paper we do not perform heavy quark expansion with respect to the charm quark, and only take the heavy quark limit of the bottom quark, then there exist two independent form factors in the $\Lambda_b \to \Lambda_c $ transition, which are denoted by $\zeta_1(q^2)$ and $\zeta_2(q^2)$. The form factors $f_i(q^2)$ and $g_i(q^2)$ can be expressed as
\begin{equation}
\begin{aligned}
&f_1=\zeta_1-\zeta_2,\,\, g_1=\zeta_1+\zeta_2,\\
&f_2=g_2=2\zeta_2, \,\,f_3=g_3=0.
\end{aligned}\label{ffrelation}
\end{equation}
 In the next section, we will estimate the power suppressed contributions from the heavy quark expansion, while the above relation still holds after including  this power correction since we have neglect the contribution from four-particle LCDAs of $\Lambda_b$. In the literature, there exists another widely uses parameterization of $\Lambda_b \to \Lambda_c$ form factors, i.e.
\begin{eqnarray}
&&\langle \Lambda_{c}(p',s')|\bar{c}\gamma_{\mu}(1-\gamma_{5})b|{\Lambda_b}(P,s)\rangle\nonumber \\
=&&\bar{u}(p',s')\left[F_{1}(q^{2})\gamma_{\mu}+F_{2}(q^2)i\sigma_{\mu\nu}\frac{q^{\nu}}{m_{\Lambda_b}}+F_{3}(q^{2})\frac{q_{\mu}}{m_{\Lambda_b}}
\right]u(P,s)\nonumber \\
 -&&\bar{u}(p',s')\left[G_{1}(q^{2})\gamma_{\mu}+G_{2}(q^2)i\sigma_{\mu\nu}\frac{q^{\nu}}{m_{\Lambda_b}}+G_{3}(q^{2})\frac{q_{\mu}}{m_{\Lambda_b}}
\right]\gamma_{5}u(P,s),\label{eq:parametrization_tradition}
\end{eqnarray}
the form factors defined above are related to $f_i,g_i$ defined in Eq.(\ref{eq:parametrization_simple}) by
\begin{eqnarray}
&&f_1=F_1-\frac{m_{\Lambda_b}+m_{\Lambda_c}}{m_{\Lambda_b}}F_2, \,\,\ \ \  f_2=F_2+F_3, \, \, \,\ \ \ \ f_3=\frac{m_{\Lambda_c}}{m_{\Lambda_b}}(F_2-F_3),\nonumber \\ 
&&g_1=G_1+\frac{m_{\Lambda_b}-m_{\Lambda_c}}{m_{\Lambda_b}}G_2,\,\,\ \ \  g_2=G_2+G_3, \, \, \,\ \ \ g_3=\frac{m_{\Lambda_c}}{m_{\Lambda_b}}(G_2-G_3),
\end{eqnarray}\label{ffrelation}
after taking the heavy bottom quark limit, the form factors $F_i$ and $G_i$ can expressed in terms of $f_i$ as follows 
\begin{eqnarray}
&&F_1=f_1+{1\over 2}(1+r_\Lambda)f_2,\,\,\,F_2={1\over 2}f_2,\nonumber \\
&&G_1=F_1,\,\,\,G_2=G_3=F_3=F_2.
\end{eqnarray}\label{ffrelation}

\subsection{Interpolating currents and correlation function}

Following the standard strategy, we start with construction of the correlation function
\begin{eqnarray}
\Pi^a_{\mu, i}(q, p')= i \int d^4 x  \, e^{i p' \cdot x} \, \langle 0 |T \{\eta^{a}(x), j_{\mu, i}(0) \}| \Lambda_b(v) \rangle \,,
\label{definition: correlator}
\end{eqnarray}
where the local current $\eta^a$ interpolates the $\Lambda_c$ and
$j_{\mu, i}$ stands for the weak transition current $\bar u \, \Gamma_{\mu, i} \, b $
with the index ``$i$" indicating a certain Lorenz structure, i.e.,
\begin{eqnarray}
j_{\mu, V} =\bar c \, \gamma_{\mu} \, b\,, &  \qquad & j_{\mu, A}=\bar c \, \gamma_{\mu}\, \gamma_5 \, b\,.
\end{eqnarray}
For the interpolation current of the $\Lambda_c$ baryon, as discussed in \cite{Khodjamirian:2011jp}, there exist the following three independent choices
\begin{eqnarray}\eta^{P}&=&\epsilon^{ijk}(u^iC\gamma_5d^j)c^k\nonumber \\
\eta^{A}&=&\epsilon^{ijk}(u^iC\gamma_5\gamma_\mu d^j)\gamma^\mu c^k\nonumber \\
\eta^{S}&=&\epsilon^{ijk}(u^iCd^j)\gamma_5c^k\end{eqnarray}
 and $i, j,$ and $k$ are the color indices and $C$ is the charge conjugation operator. The correlation function will vanish if the S-type current is employed, thus we only adopt the P-type and A-type operators in our study.
The  coupling of $\Lambda_c$ as well as its party odd partner with the interpolating current $\eta^a$ (the decay constant),    is defined as
\begin{eqnarray}\langle 0|\eta^a|\Lambda_c\rangle&=&m_{\Lambda_c}\lambda^a_{\Lambda_c}u(p')\nonumber \\
\langle 0|\eta^a|\Lambda^\ast_c\rangle&=&m_{\Lambda^\ast_c}\lambda^a_{\Lambda^\ast_c}\gamma_5u(p')\end{eqnarray}
At hadronic level, the correlation function can be expressed in terms of the matrix elements of the currents sandwiched by the hadronic states
\begin{eqnarray}
\Pi^a_{ \mu,i}=&& \frac{1}{m_{\Lambda_c}-p^{\prime 2}}\sum_{s'}\left\langle 0\left|\eta^{a}\right| \Lambda_c\left(p^{\prime},s'\right)\right\rangle\left\langle \Lambda_c\left(p^{\prime},s'\right)\left| j_{\mu, i}(0)\right| \Lambda_b(v,s)\right\rangle \nonumber\\
+&&\frac{1}{m_{\Lambda^\ast_c}^{2}-p^{\prime 2}}
\sum_{s'}\langle 0|\eta^a| \Lambda^\ast_c(p^{\prime},s')\rangle \langle \Lambda^\ast_c(p^{\prime},s')|j_{\mu,i}(0)| \Lambda_b(v,s)\rangle\nonumber\\
+&&{1\over \pi} \int_{s_0}^\infty {ds\over s-p'^2-i0}\left[\rho_{1i}^h(s)\gamma_\mu+\rho_{2i}^h(s)v_\mu+\rho_{3i}^h(s)v'_\mu\right ]\Lambda_b(v),
\end{eqnarray}
where $\rho^h_{ai}(s)$ denoting the hadronic spectral densities of all excited and continuum states with the quantum numbers of $\Lambda_c$ and $\Lambda_c^\ast$. It is then a straightforward task to write down the hadronic representations for the correlation functions
defined with various weak currents. For the vector current, we have

\begin{eqnarray}
\Pi^a_{\mu, V}(p, q) &=& \frac{m_{\Lambda_c}\lambda^a_{\Lambda_c} }
{m_{\Lambda_c}^2 -p^{\prime 2} } \left(\not\!p^{\prime}+m_{\Lambda_c}\right)\left[\gamma_{\mu}f_{1}(q^{2})+v_{\mu}f_{2}(q^{2})
+{v'_{\mu}}f_{3}(q^{2})
\right] u\left(v, s\right)  \nonumber
 \\&+& \frac{m_{\Lambda_c^\ast}\lambda^a_{\Lambda_c^\ast} }
{m_{\Lambda_c^\ast}^2 -p^{\prime 2} } \left(-\not\!p^{\prime}+m_{\Lambda_c^\ast}\right)\left[\gamma_{\mu}\tilde f_{1}(q^{2})+v_{\mu}\tilde f_{2}(q^{2})
+{v'_{\mu}}\tilde f_{3}(q^{2})\right] u\left(v, s\right)+...
\end{eqnarray}
For the  $\Pi^a_{\mu, A}(p', q)$, only the replacement $f_i\to g_i, \tilde f_i \to \tilde g_i$ is required.
%\begin{eqnarray}
%\Pi^a_{\mu, A}(p', q) &=& \frac{m_{\Lambda_c}\lambda^a_{\Lambda_c} }
%{m_{\Lambda_c}^2 -p^{\prime 2} } \left(\not\!p^{\prime}+m_{\Lambda_c}\right)\left[v_{\mu}G_{1}(q^{2})
%+\frac{p'_{\mu}}{m_{\Lambda_c}}G_{2}(q^{2})
%+\gamma_{\mu}G_{3}(q^{2})\right] \gamma_5 u\left(v, s\right)  \nonumber
% \\&+& \frac{m_{\Lambda_c^\ast}\lambda^a_{\Lambda_c^\ast} }
%{m_{\Lambda_c^\ast}^2 -p^{\prime 2} } \gamma_5 %\left(\not\!p^{\prime}+m^\ast_{\Lambda_c}\right)\left[v_{\mu}\tilde G_{1}(q^{2})
%+\frac{p'_{\mu}}{m_{\Lambda_c^\ast}}\tilde G_{2}(q^{2})
%+\gamma_{\mu}\tilde G_{3}(q^{2})\right] \gamma_5 u\left(v, s\right)+...
%\end{eqnarray}
 Through the analysis of the Lorentz structures, the correlation function can be parameterized as
\begin{eqnarray}
\Pi^a_{ \mu, V}(p', q) &=& \left[\Pi^a_{1+}\gamma_\mu+\Pi^a_{1-}\not\! p^{\prime}\gamma_\mu+\Pi^a_{2+}v_\mu
+\Pi^a_{2-} v_\mu\not\!p^{\prime}{+}\Pi^a_{3+}v'_\mu {+}\Pi^a_{3-}\not\!p^{\prime}v'_\mu \right]u\left(v\right)+...,
\end{eqnarray}
\begin{eqnarray}
\Pi^a_{ \mu, A}(p', q) &=& \left[\overline\Pi^a_{1+}\gamma_\mu+\overline\Pi^a_{1-}\not\! p^{\prime}\gamma_\mu+\overline\Pi^a_{2+}v_\mu
+\overline\Pi^a_{2-}v_\mu\not\!p^{\prime}{+}\overline\Pi^a_{3+}v'_\mu {+}\overline\Pi^a_{3-}\not\! p^{\prime}v'_\mu \right]\gamma_5u\left(v\right)+...,
\end{eqnarray}
then the scalar correlation functions can be expressed in  terms of the form factors as follows
\begin{eqnarray}
\Pi_{i+}^a &=&\frac{m_{\Lambda_c}^2\lambda^a_{\Lambda_c} }
{m_{\Lambda_c}^2 -p^{\prime 2} }f_i +\frac{m_{\Lambda_c^\ast}^2\lambda^a_{\Lambda_c^\ast} }
{m_{\Lambda_c^\ast}^2 -p^{\prime 2} }\tilde f_i+...\nonumber \\
\Pi_{i-}^a &=&\frac{m_{\Lambda_c}\lambda^a_{\Lambda_c} }
{m_{\Lambda_c}^2 -p^{\prime 2} }f_i -\frac{m_{\Lambda_c^\ast}\lambda^a_{\Lambda_c^\ast} }
{m_{\Lambda_c^\ast}^2 -p^{\prime 2} }\tilde f_i+...
%\Pi_2^a &=&\frac{m_{\Lambda_c}\lambda^a_{\Lambda_c} }
%{m_{\Lambda_c}^2 -p^{\prime 2} }m_{\Lambda_c}F_1 -\frac{m_{\Lambda_c^\ast}\lambda^a_{\Lambda_c^\ast} }
%{m_{\Lambda_c^\ast}^2 -p^{\prime 2} }m_{\Lambda_c^\ast}\tilde F_1\nonumber \\
%\Pi_3^a &=&\frac{m_{\Lambda_c}\lambda^a_{\Lambda_c} }
%{m_{\Lambda_c}^2 -p^{\prime 2} }m_{\Lambda_c}F_2 +\frac{m_{\Lambda_c^\ast}\lambda^a_{\Lambda_c^\ast} }
%{m_{\Lambda_c^\ast}^2 -p^{\prime 2} }m_{\Lambda_c^\ast}\tilde F_2\nonumber \\
%\Pi_4^a &=&\frac{m_{\Lambda_c}\lambda^a_{\Lambda_c} }
%{m_{\Lambda_c}^2 -p^{\prime 2} }m_{\Lambda_c}F_2 -\frac{m_{\Lambda_c^\ast}\lambda^a_{\Lambda_c^\ast} }
%{m_{\Lambda_c^\ast}^2 -p^{\prime 2} }m_{\Lambda_c^\ast}\tilde F_2\nonumber \\
%\Pi_5^a &=&\frac{m_{\Lambda_c}\lambda^a_{\Lambda_c} }
%{m_{\Lambda_c}^2 -p^{\prime 2} }m_{\Lambda_c}F_3 -\frac{m_{\Lambda_c^\ast}\lambda^a_{\Lambda_c^\ast} }
%{m_{\Lambda_c^\ast}^2 -p^{\prime 2} }m_{\Lambda_c^\ast}\tilde F_3\nonumber \\
%\Pi_6^a &=&\frac{m_{\Lambda_c}\lambda^a_{\Lambda_c} }
%{m_{\Lambda_c}^2 -p^{\prime 2} }m_{\Lambda_c}F_3 +\frac{m_{\Lambda_c^\ast}\lambda^a_{\Lambda_c^\ast} }
%{m_{\Lambda_c^\ast}^2 -p^{\prime 2} }m_{\Lambda_c^\ast}\tilde F_3\nonumber \\
\label{scalar correlator}\end{eqnarray}
For the correlation function with axial vector part of the weak current, $\Pi_i \to\overline \Pi_i$, the replacement $f_i \to g_i$, $\tilde f_i \to \tilde g_i$ is needed in the Eq. (\ref{scalar correlator}).

\subsection{Tree-level LCSR}

Now, we turn to compute the correlation function $\Pi_{i\mu, a}(p, q)$
with space-like interpolating momentum with $|\bar n \cdot p| \sim {\cal O} (\Lambda)$
and $n \cdot p\sim m_{\Lambda_b}$ at partonic level. The correlation function can be factorized into the convolution of the hard kernel and the LCDAs of $\Lambda_b$-baryon, i.e.
\begin{eqnarray}
\Pi^{{\rm LP},a}_{i\mu, \gamma}(p', q)=\int d \omega_1^{\prime} \int d \omega_2^{\prime} \,
T^{(a,0)}_{i\mu,\alpha \beta \gamma \delta}( p^{\prime}, q, \omega_1^{\prime}, \omega_2^{\prime})\,
\Phi_{\Lambda_b}^{ \, \alpha \beta \delta}(\omega_1^{\prime}, \omega_2^{\prime}) \,,
\label{tree-level factorization at partonic level}
\end{eqnarray}
where the definition of the most general light-cone hadronic matrix element
in coordinate space \cite{Bell:2013tfa} is given by
\begin{eqnarray}
\Phi_{\Lambda_b}^{\alpha \beta \delta}(t_1, t_2)
&\equiv&  \epsilon_{i j k} \, \langle 0 | \left [u^{\rm T}_{i} (t_1 \bar n) \right ]_{\alpha} \,
[0, t_1 \bar n] \, \left [d_{j} (t_2 \bar n) \right ]_{\beta} \, [0, t_2 \bar n] \,
\left [ b_{k}(0)\right ]_{\delta} | \Lambda_b(v) \rangle \nonumber \\
&=&  \frac{1}{4 } \, \left \{ f_{\Lambda_b}^{(1)}(\mu) \,
\left [ \tilde{M}_1(v, t_1, t_2) \, \gamma_5 \, C^{T} \right ]_{\beta \alpha}
+  f_{\Lambda_b}^{(2)}(\mu) \,
\left [ \tilde{M}_2(v, t_1, t_2) \, \gamma_5 \, C^{T} \right ]_{\beta \alpha} \right \} \,
\left [ \Lambda_b(v) \right ]_{\delta}   \,. \nonumber \\
\end{eqnarray}
Performing the Fourier transformation and including the next-to-leading order terms off the light-cone leads to
the momentum space light-cone projector in $D$ dimensions
\begin{eqnarray}
M_2(\omega_1^{\prime},\omega_2^{\prime}) &=& \frac {\! \not n}{2} \, \psi_2(\omega_1^{\prime},\omega_2^{\prime})
+ \frac {\! \not \bar  n}{2} \, \psi_4(\omega_1^{\prime},\omega_2^{\prime}) \nonumber
\\
&& -\frac{1}{D-2} \, \gamma_{\perp}^{\mu} \, \left  [ \psi_{\perp, 1}^{+-}(\omega_1^{\prime},\omega_2^{\prime}) \,
\frac{\! \not n  \, \! \not \bar  n}{4}  \, \frac{\partial}{\partial k_{1 \perp}^{\mu}}
+ \psi_{\perp, 1}^{-+}(\omega_1^{\prime},\omega_2^{\prime}) \,
\frac{\! \not \bar n  \, \! \not  n}{4}  \, \frac{\partial}{\partial k_{1 \perp}^{\mu}} \right  ] \nonumber
\\
&& -\frac{1}{D-2} \, \gamma_{\perp}^{\mu} \, \left [{ \psi_{\perp, 2}^{+-}(\omega_1^{\prime},\omega_2^{\prime})} \,
\frac{\! \not n  \, \! \not \bar  n}{4}  \, \frac{\partial}{\partial k_{2 \perp}^{\mu}}
+ \psi_{\perp, 2}^{-+}(\omega_1^{\prime},\omega_2^{\prime}) \,
\frac{\! \not \bar n  \, \! \not  n}{4}  \, \frac{\partial}{\partial k_{2 \perp}^{\mu}} \right  ]  \,,
\label{chiral-odd projector}\\
M_1(\omega_1^{\prime},\omega_2^{\prime}) &=& \frac{\! \not \bar  n \, \! \not  n }{8} \,
\psi_{3}^{+-}(\omega_1^{\prime},\omega_2^{\prime})
+ \frac{\! \not  n  \, \! \not \bar  n}{8} \,
\psi_{3}^{-+}(\omega_1^{\prime},\omega_2^{\prime}) \nonumber
\\
&& - \frac{1}{D-2} \left [ \psi_{\perp, 3}^{(1)}(\omega_1^{\prime},\omega_2^{\prime}) \! \not  v
\, \gamma_{\perp}^{\mu}  \, \frac{\partial}{\partial k_{1 \perp}^{\mu}}
+ \psi_{\perp, 3}^{(2)}(\omega_1^{\prime},\omega_2^{\prime}) \, \gamma_{\perp}^{\mu}  \,
\! \not  v \, \frac{\partial}{\partial k_{2 \perp}^{\mu}}  \right ] \nonumber
\\
&& - \frac{1}{D-2} \left [ \psi_{\perp, Y}^{(1)}(\omega_1^{\prime},\omega_2^{\prime}) \! \not \bar n
\, \gamma_{\perp}^{\mu}  \, \frac{\partial}{\partial k_{1 \perp}^{\mu}}
+ \psi_{\perp, Y}^{(2)}(\omega_1^{\prime},\omega_2^{\prime}) \,  \gamma_{\perp}^{\mu}  \,
\! \not  \bar n \, \frac{\partial}{\partial k_{2 \perp}^{\mu}}  \right ] \,,
\label{chiral-even projector}
\end{eqnarray}
where we have adjusted the notation of the $\Lambda_b$-baryon DA defined in \cite{Bell:2013tfa}.
Applying the equations of motion in the Wandzura-Wilczek approximation  yields
\begin{eqnarray}
\psi_{\perp, 1}^{-+}(\omega_1^{\prime},\omega_2^{\prime})=\omega_1^{\prime} \,
\psi_4(\omega_1^{\prime},\omega_2^{\prime}) \,, \qquad
\psi_{\perp, 2}^{+-}(\omega_1^{\prime},\omega_2^{\prime})=\omega_2^{\prime} \,
\psi_4(\omega_1^{\prime},\omega_2^{\prime})  \,.
\label{EOM for Lambdab DA}\end{eqnarray}
%%%%%%%%%%%
\begin{figure}
\begin{center}
\includegraphics[width=0.4 \columnwidth]{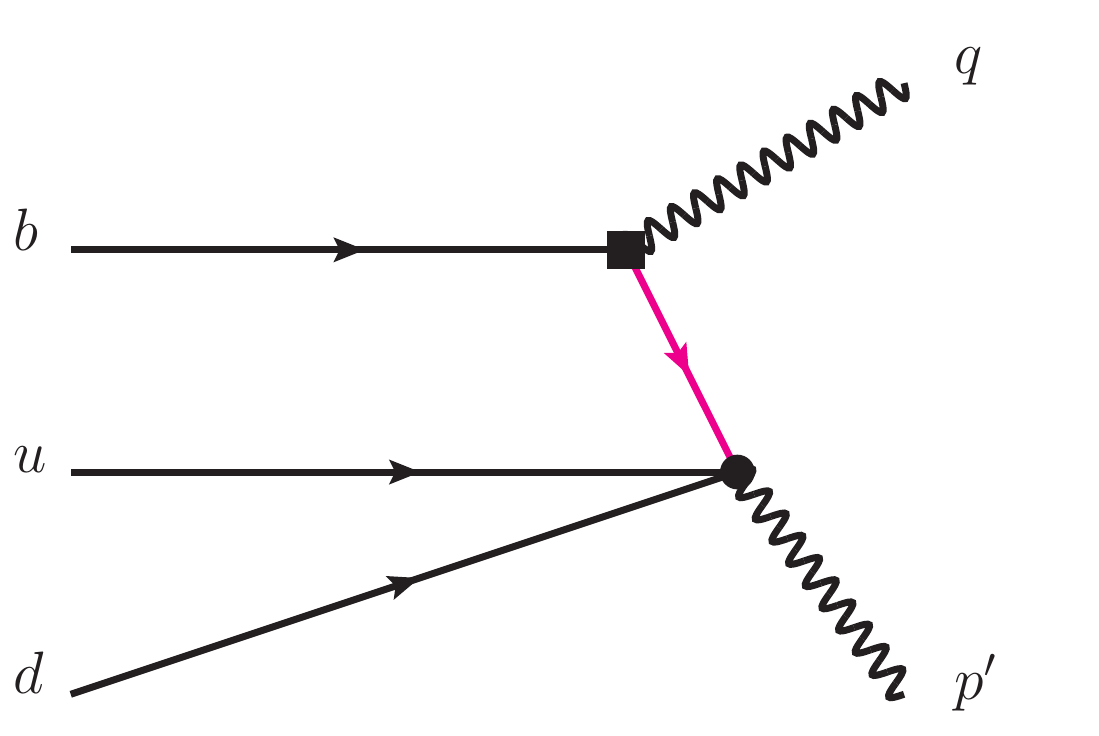}
\vspace*{0.1cm}
\caption{Diagrammatical representation of the correlation function
$\Pi_{\mu, a}(n \cdot p^{\prime},\bar n \cdot p^{\prime})$ at tree level,
where the black square denotes the weak transition vertex, the black blob represents
the Dirac structure of the $\Lambda_c$-baryon current and the pink internal line indicates the 
propagator of the charm quark. }
\label{fig:tree_correlator}
\end{center}
\end{figure}
%%%%%%%%%%%
Evaluating the diagram in Fig. \ref{fig:tree_correlator} leads to the leading-order hard kernel
\begin{eqnarray}
T^{(\rm P,0)}_{\alpha \beta \gamma \delta}(p,q)&=&-\left[C\gamma^5\right]_{\alpha\beta} \left[{\not\! p-\not\! k+m_c\over (p-k)^2-m_c^2}\gamma_\mu(I,\gamma_5)\right]_{\gamma\delta}\nonumber \\
T^{(\rm A,0)}_{\alpha \beta \gamma \delta}(p,q)&=&-\left[C\gamma^5\gamma^\rho\right]_{\alpha\beta} \left[\gamma_\rho{\not\! p-\not\! k+m_c\over (p-k)^2-m_c^2}\gamma_\mu(I,\gamma_5)\right]_{\gamma\delta}
\label{tree-level factorization at partonic level}.
\end{eqnarray}
where $k=k_1+k_2$ with $k_{1,2}$ standing for the momentum of the two soft light quarks inside $\Lambda_b$-baryon. Inserting the hard functions and the DAs into the correlation functions, we can arrive at the partonic expression of the correlation functions. We note that in order to match the light-like vector $n$ and $\bar n$ in the definition of the DAs of $\Lambda_b$ baryon and the momentum $p',q$ in the parametrization of the correlation function,  we need to perform the replacement
\begin{eqnarray}
\psi(\omega)\not\! \bar n\to \overline{\psi}(\omega)\not\!\partial_{p'},\,\, \psi(\omega)\not\!  n\to 2\psi(\omega)\not\! v -\overline{\psi}(\omega)\not\!\partial_{p'}\,\,{\rm or}\,\,\overline{\psi}(\omega)(2\not\! vv\cdot \partial_{p'}-\not\!\partial_{p'}),
\end{eqnarray}
where $\bar \psi(\omega)=\int_0^\omega \eta\psi(\eta)d\eta$. The obtained invariant amplitudes $\Pi^{a}_{i}$ can be expressed by the following dispersion integral
\begin{eqnarray}
\Pi^{a}_{i}(p^2 , q^2)=\frac{1}{\pi} \int^{\infty}_0 \frac{ds}{s-p^2} {\rm Im}_s \Pi^a_i(s,q^2).
\end{eqnarray}
Taking advantage of the quark hadron duality ansatz, namely,  equalizing the contributions from the continuum states and higher states  in the hadronic expression and the dispersion integral with the  lower limit being the threshold $s_0$ in the partonic expression of the correlation function, and performing the Borel transform, we can obtain the sum rules at leading power. For the P-type current, the sum rules of the form factors can be written by
\begin{eqnarray}
f_i^{\rm P,LP}&=&-{m_{\Lambda_b}^2\over m_{\Lambda_c}(m_{\Lambda_c}+m_{\Lambda_c^\ast})\lambda^{\rm P}_{\Lambda_c^\ast}}\int_0^1du\int_0^{\sigma_0}{\sigma d\sigma\over \bar\sigma}[\rho_{i+,a}^{\rm P}(\sigma)+m_{\Lambda_c^\ast}\rho_{i-,a}^{\rm P}(\sigma)]e^{(m_{\Lambda_c}^2-s(\sigma))/M_B^2}
%\tilde F_i^{\rm P,LP}&=&-{m_{\Lambda_b}^2\over m_{\Lambda_c^\ast}(m_{\Lambda_c}+m_{\Lambda_c})\lambda^P_{\Lambda_c^\ast}}\int_0^1du\int_0^{\sigma_0}{\sigma d\sigma\over \bar\sigma}[\rho_{i+,a}^{\rm P}(\sigma)-m_{\Lambda_c}\rho_{i-,a}^{\rm P}(\sigma)]e^{(m_{\Lambda_c^\ast}^2-s(\sigma))/M_B^2}
\end{eqnarray}
where the nonzero spectrum densities read
\begin{eqnarray}
{\rho^{\rm P}_{1+,a}}&=&-{1\over 4}f_{\Lambda_b}^{(1)}\left(\psi_3^{+-}+ \psi_3^{-+}\right)(m_{\Lambda_b}\sigma+m_c),\,\,\rho^{\rm P}_{2+,a}={1\over 2}f_{\Lambda_b}^{(1)}\left(\psi_3^{+-}+ \psi_3^{-+}\right)m_{\Lambda_b}\sigma,\nonumber \\
\rho^{\rm P}_{2-,a}&=&{-{1\over 4}f_{\Lambda_b}^{(1)}\left(\psi_3^{+-}+ \psi_3^{-+}\right)}. 
\end{eqnarray}
%%%%%%%%%%%%%%%%%%%%%%%%%%%%%%%%%%%%%%%%%%%%%%%%%%%%%%%%%%%%%%%%%%%%%%%%%%%%%%%%%%%%
 For A-type current, we have
\begin{eqnarray}
f_i^{\rm A, LP}&=&{-e^{m_{\Lambda_c}^2/M_B^2}\over m_{\Lambda_c}(m_{\Lambda_c}+m_{\Lambda_c^\ast})\lambda^A_{\Lambda_c^\ast}}\int_0^1du\bigg\{\int_0^{\sigma_0}{m_{\Lambda_b}^2\sigma d\sigma\over \bar\sigma}[\rho_{i+,a}^{A}(\sigma)+m_{\Lambda_c^\ast}\rho_{i-,a}^{A}(\sigma)]e^{-s(\sigma)/M_B^2}
\nonumber \\&+&{[\rho_{i+,b}^{A}(\sigma_0)+m_{\Lambda_c^\ast}\rho_{i-,b}^{A}(\sigma_0)]\over {m_{\Lambda_b}\bar \sigma_0}^2}\eta(\sigma_0)e^{-s_0/M_B^2}+\int_0^{\sigma_0}{m_{\Lambda_b} d\sigma\over {\bar\sigma}^2}{[\rho_{i+,b}^{A}(\sigma)+m_{\Lambda_c^\ast}\rho_{i-,b}^{A}(\sigma)]\over M_B^2}e^{-s(\sigma)/M_B^2}\bigg\}\nonumber \\
%\tilde F_i^A&=&-{e^{m_{\Lambda_c^\ast}^2/M_B^2}\over m_{\Lambda_c}(m_{\Lambda_c}+m_{\Lambda_c^\ast})\lambda^A_{\Lambda_c}}\int_0^1du\bigg\{\int_0^{\sigma_0}{m_{\Lambda_b}^2\sigma d\sigma\over \bar\sigma}[\rho_{i+,a}^{\textcolor{red}{A}}(\sigma)-m_{\Lambda_c}\rho_{i-,a}^{\textcolor{red}{A}}(\sigma)]e^{-s(\sigma)/M_B^2}
%\nonumber \\&+&{[\rho_{i+,\textcolor{red}{b}}^{\textcolor{red}{A}}(\sigma_0)-m_{\Lambda_c^\ast}\rho_{i-,\textcolor{red}{b}}^{\textcolor{red}{A}}(\sigma_0)]\over {m_{\Lambda_b}\bar \sigma_0}^2}\eta(\sigma_0)e^{-s_0/M_B^2}+\int_0^{\sigma_0}{m_{\Lambda_b} d\sigma\over {\bar\sigma}^2}{[\rho_{i+,b}^{\textcolor{red}{A}}(\sigma)-m_{\Lambda_c}\rho_{i-,b}^{\textcolor{red}{A}}(\sigma)]\over M_B^2}e^{-s(\sigma)/M_B^2}\bigg\}\nonumber \\
\end{eqnarray}
where the nonzero spectrum densities are given below
\begin{eqnarray}
\rho^{\rm A}_{1+,a}&=&-{1\over 2}f_{\Lambda_b}^{(2)}{[2(\bar \psi_4-\bar \psi_2)}+2\psi_2(2m_{\Lambda_c}v\cdot v'-m_{\Lambda_b}\sigma-m_c)+(\psi_{\perp, 1}+\psi_{\perp, 2})],\,\,
\rho^{\rm A}_{1-,a}={-}f_{\Lambda_b}^{(2)}{\psi_2}\nonumber \\
\rho^{\rm A}_{1+,b}&=&f_{\Lambda_b}^{(2)}(\bar \psi_4-\bar \psi_2)m_c(m_{\Lambda_b}\sigma+m_c),\,\,
\rho^{\rm A}_{1-,b}=f_{\Lambda_b}^{(2)}{m_c}(\bar \psi_4-\bar \psi_2)\\
\rho^{\rm A}_{2+,a}&=&-{2}f_{\Lambda_b}^{(2)}m_c\psi_2,\,\,
\rho^{\rm A}_{2-,a}={2}f_{\Lambda_b}^{(2)}\psi_2,
\nonumber \\
\rho^{\rm A}_{2+,b}&=&-{2}f_{\Lambda_b}^{(2)}(\bar \psi_4-\bar \psi_2)m_cm_{\Lambda_b}\sigma
\nonumber 
\end{eqnarray}
The form factors $g_i$ can be obtained from $f_i$ directly so that we do not present the explicit expressions. 

\subsection{Power suppressed contribution from heavy quark expansion}
Now, we discuss the power suppressed contribution from heavy quark expansion, to achieve the target, we should replace the leading power heavy quark field in the heavy-to-light current by the NLP suppressed one in the QCD calculation, i.e.
\begin{eqnarray}
\bar c\Gamma h_v \to \bar c \Gamma {i\not\! D\over 2m_b}h_v.
\end{eqnarray}
Then the correlation function(we take the correlation function with P-type interpolation current as an example) turns to
\begin{eqnarray}
\Pi^{\rm P,NLP}_{\mu, a}(q, p')= i \int d^4 x  \, e^{i p' \cdot x} \, \langle 0 |T \{\epsilon^{ijk}(u^iC\gamma_5d^j)c^k(x), \bar c(0) \Gamma_a {i\not\! D\over 2m_b}h_v(0)  \}| \Lambda_b(v) \rangle \,.
\end{eqnarray}
Contracting the charm quark field, we have
\begin{eqnarray}
\Pi^{\rm P,NLP}_{\mu, a}(q, p')= i \int{d^4k\over (2\pi)^4}\int d^4 x  \, {ie^{i (p'-k) \cdot x}\over k^2-m_c^2} \, \langle 0 |\epsilon^{ijk}(u^iC\gamma_5d^j)(\not\! k+m_c) \Gamma_a {i\not\! D\over 2m_b}h_v(0) | \Lambda_b(v) \rangle \,,
\end{eqnarray}
The QCD equation of motion indicates that
\begin{eqnarray}
[q_1(x)\Gamma' q_2(x)] \, \Gamma \, {\overrightarrow{D}}_{\rho} \,  h_v(0)
&=& \partial_{\rho} \left ( [q_1(x)\Gamma' q_2(x)] \, \Gamma \, h_v(0) \right ) \,
\nonumber \\ &+& \, i \, \int_0^1 d u \, \bar u \,\, [q_1(x)\Gamma' q_2(x)] \,g_s \, G_{\lambda \rho}(u x) \,
x^{\lambda} \, \Gamma \,  h_v(0) \nonumber \\
& - &\, {\partial \over \partial x^{\rho}}[q_1(x)\Gamma' q_2(x)] \, \Gamma \, h_v(0)  \,.
\end{eqnarray}
The matrix element of the second term results in the convolution of the hard function and the four-point LCDA of the $\Lambda_b$-baryon, which has not been studied yet,  thus we leave this part for the future study. In addition, the derivative on the gauge link will also result in an additional gluon field, which will also be neglected in the present study.  Then the correlation function reads
\begin{eqnarray}
\Pi^{\rm P,NLP}_{\mu, a}(q, p')&\simeq& -i {1\over 2m_b}\int{d^4k\over (2\pi)^4}\int d^4 x  \, {e^{i (p'-k) \cdot x}\over k^2-m_c^2} \, \partial_\rho \langle 0 |\epsilon^{ijk}(u^iC\gamma_5d^j)(\not\! k+m_c) \Gamma_a \gamma^\rho h_v(0) | \Lambda_b(v) \rangle \,\nonumber \\
&+&i {1\over 2m_b}\int{d^4k\over (2\pi)^4}\int d^4 x  \, {e^{i (p'-k) \cdot x}\over k^2-m_c^2} \, {\partial\over \partial x^\rho} \langle 0 |\epsilon^{ijk}(u^iC\gamma_5d^j)(\not\! k+m_c) \Gamma_a \gamma^\rho h_v(0) | \Lambda_b(v) \rangle ,
\end{eqnarray}
The first term can be evaluated directly. Taking advantage of the definition of the heavy quark field in HQET, the partial derivative leads to a simple nonperturative parameter
\begin{eqnarray}
 \partial_\rho\langle 0 |\epsilon^{ijk}(u^iC\gamma_5d^j)(\not\! k+m_c) \Gamma_a \gamma^\rho h_v(0) | \Lambda_b(v) \rangle =-i\bar \Lambda \langle 0 |\epsilon^{ijk}(u^iC\gamma_5d^j)(\not\! k+m_c) \Gamma_a h_v(0) | \Lambda_b(v) \rangle\, ,
\end{eqnarray}
where the nonperturavtive parameter is regarded to be the mass different between the $\Lambda_b$-baryon and the $b$-quark for a good approximation, i.e., $\bar \Lambda\simeq m_{\Lambda_b} - m_b$. For the second term,  performing the integration by part yields additional $\omega=v\cdot (p'-k)$ in the integrand. Combine this two parts together, we have
\begin{eqnarray}
\Pi^{\rm P,NLP}_{\mu, a}(q, p')&=& -{\bar \Lambda\over 2m_b}\Pi^{\rm P,LP}_{\mu, a}(q, p')-{1\over 2m_b}\int \omega' d \omega^{\prime} \int_0^1 du \,
T^{\rm (P,0)}_{\alpha \beta \gamma \delta}( p^{\prime}, q, u\omega^{\prime}, \bar u\omega^{\prime})\,
\Phi_{\Lambda_b}^{ \, \alpha \beta \delta}(\omega^{\prime}, u).
\end{eqnarray}
Finally, we arrive at the sum rules of the form factors at NLP and they are written by
\begin{eqnarray}
f_i^{\rm P,NLP}&=&{m_{\Lambda_b}^2\over m_{\Lambda_c}(m_{\Lambda_c}+m_{\Lambda_c^\ast})\lambda^P_{\Lambda_c}}\int_0^1du\int_0^{\sigma_0}{\bar \Lambda+\sigma m_{\Lambda_b}\over 2m_b}{\sigma d\sigma\over \bar\sigma}[\rho_{i+,a}^P(\sigma)+m_{\Lambda_c}\rho_{i-,a}^P(\sigma)]e^{(m_{\Lambda_c}^2-s(\sigma))/M_B^2}\nonumber ,\\
%\tilde F_i^{P,NLP}&=&-{m_{\Lambda_b}^2\over m_{\Lambda_c^\ast}(m_{\Lambda_c}+m_{\Lambda_c^\ast})\lambda^P_{\Lambda_c^\ast}}\int_0^1du\int_0^{\sigma_0}{-\bar \Lambda-\sigma m_{\Lambda_b}\over 2m_b}{\sigma d\sigma\over \bar\sigma}[\rho_{i+,a}^P(\sigma)-m_{\Lambda_c^\ast}\rho_{i-,a}^P(\sigma)]e^{(m_{\Lambda_c^\ast}^2-s(\sigma))/M_B^2} .\nonumber \\
\end{eqnarray}
From the above result, we can see that the power suppressed contribution considered in the present work is to add a factor $(\Lambda-\sigma m_{\Lambda_b})/( 2m_b)$ in the integrand of the leading power contirbuiton if P-type interpolation current is employed. For the A-type interpolation current, we need to perform  a more complicated modification since there exist $\bar \psi_i(\omega)$ in the integrand. The specific operation is as follows: in the spectrum density $\rho_{i\pm,a(b)}^A$, we multiply   $(\Lambda-\sigma m_{\Lambda_b})/( 2m_b)$ to the terms proportional to $\psi_i(\sigma)$, and replace $\bar \psi_i$ by $\tilde{\psi}_i/m_b$, where  $\tilde{\psi}_i(\omega)$ is defined by $\tilde{\psi}_i(\omega)=\int_0^\omega\eta^2\psi_i(\eta)d\eta$.

\section{Numerical analysis}
DAs of the $\Lambda_b$ baryon are the fundamental ingredients for the LCSR of the form factors considered in the present paper, but they are not well established so far due to our poor understanding of QCD dynamics inside the heavy baryon system. In \cite{Ball:2008fw,Ali:2012pn,Bell:2013tfa} several different models of the LCDAs for the $\Lambda_b$ baryon have been suggested up to twist-4(not including the twist of the heavy quark field), we consider the following three different models. The first one is obtained from the calculation with QCDSR \cite{Ball:2008fw}, thus it is named as the QCDSR-model. The specific form for the LCDAs $\psi_2(\omega, u),\psi_3^{+-}(\omega, u),\psi_3^{-+}(\omega, u),\psi_4(\omega, u)$ reads
\begin{eqnarray}
   \psi_2(\omega, u)&=&\frac{15}{2\mathcal{N}}\omega^2 u(1-u)\int^{s^{\Lambda_b}_0}_{\omega/2}ds\ e^{-s/\tau} (s-\omega/2) \nonumber \\
   \psi^{+-}_{3}(\omega,u)&=&\frac{15}{\mathcal{N}}\omega u\int^{s^{\Lambda_b}_0}_{\omega/2}ds\ e^{-s/\tau}(s-\omega/2)^2 \nonumber \\
   \psi^{-+}_{3}(\omega,u)&=&\frac{15}{\mathcal{N}}\omega (1-u)\int^{s^{\Lambda_b}_0}_{\omega/2}ds\ e^{-s/\tau}(s-\omega/2)^2 \nonumber \\
   \psi_4(\omega,u)&=&\frac{5}{\mathcal{N}}\int^{s^{\Lambda_b}_0}_{\omega/2}ds\ e^{-s/\tau}(s-\omega/2)^3
\end{eqnarray}
with
$\mathcal{N}=\int^{s^{\Lambda_b}_0}_0ds\ s^5e^{-s/\tau}$. $\tau$ is the Borel parameter which is constrained in the interval $0.4<\tau<0.8$ GeV and ${s^{\Lambda_b}_0}=1.2$ GeV is the continuum threshold. The other two phenomenological models are proposed in \cite{Bell:2013tfa}, and they are called Exponential-model and Free parton-model respectively. For the  Exponential-model,
\begin{eqnarray}
    \psi_2(\omega,u)&=&\frac{\omega^2 u(1-u)}{\omega^4_0}e^{-\omega/\omega_0} \nonumber \\
    \psi^{+-}_3(\omega, u)&=&\frac{2\omega u}{\omega^3_0}e^{-\omega/\omega_0} \nonumber \\
    \psi^{-+}_3(\omega, u)&=&\frac{2\omega (1-u)}{\omega^3_0}e^{-\omega/\omega_0} \nonumber \\
        \psi_4(\omega, u)&=&\frac{1}{\omega^2_0}e^{-\omega/\omega_0}
\end{eqnarray}
where $\omega_0=0.4\pm 0.1$ Gev measures the average of the two light quarks inside the $\Lambda_b$ baryon. The DAs in the free-parton model take the following form
\begin{eqnarray}
    \psi_2(\omega, u)&=&\frac{15\omega^2 u(1-u)(2\bar{\Lambda}-\omega)}{4\bar{\Lambda}^5}\theta(2\bar{\Lambda}-\omega) \nonumber \\
    \psi^{+-}_3(\omega, u)&=&\frac{15\omega u (2\bar{\Lambda}-\omega)^2}{4\bar{\Lambda}^5}\theta(2\bar{\Lambda}-\omega) \nonumber \\
    \psi^{-+}_3(\omega, u)&=&\frac{15\omega (1-u) (2\bar{\Lambda}-\omega)^2}{4\bar{\Lambda}^5}\theta(2\bar{\Lambda}-\omega) \nonumber \\
    \psi_4(\omega, u)&=&\frac{5(2\bar{\Lambda}-\omega)^3}{8\bar{\Lambda}^5}\theta(2\bar{\Lambda}-\omega)
\end{eqnarray}
where $\theta(2\bar{\Lambda}-\omega)$ is the step-function, and $\bar{\Lambda}=m_{\Lambda_b}-m_b\approx 1\pm0.2$ GeV.
The first-order terms off the light-cone is not significant numerically,  while they are required to guarantee the gauge invariance. In this work, the DAs of these terms are given by :
\begin{eqnarray}
  \psi^{+-}_{\perp, 1}(\omega, u)&=&  \psi^{-+}_{\perp, 2}=\psi^{(1)}_{\perp, 3}=\psi^{(2)}_{\perp, 3}=\frac{\omega^2 u(1-u)}{\omega_0^3} e^{-\omega/\omega_0} \nonumber \\
  \psi^{-+}_{\perp, 1}(\omega, u)&=&\frac{\omega u}{\omega_0^2} e^{-\omega/\omega_0} \nonumber \\
  \psi^{+-}_{\perp, 2}(\omega, u)&=&\frac{\omega (1-u)}{\omega_0^2} e^{-\omega/\omega_0} \nonumber \\
  \psi^{(1)}_{\perp, Y}(\omega, u)&=&\frac{\omega u(\omega_0-\omega(1-u))}{2\omega_0^3}e^{-\omega/\omega_0}\nonumber \\
  \psi^{(2)}_{\perp, Y}(\omega, u)&=&\frac{\omega (1-u)(\omega_0-\omega u)}{2\omega_0^3}e^{-\omega/\omega_0}
\end{eqnarray}
where $\omega_0=0.4\pm0.1$ GeV. The numerical values of the other parameters, such as the masses of the corresponding baryons, the quark masses, the coupling parameters of the baryons, the Borel mass, the threshold parameters are collected in Table. \ref{Tab:input}. In this table, we use $\overline{MS}$ mass for charm quark which appears in the partonic evaluation of the correlation functions. For the bottom quark mass, we take advantage of the potential subtraction(PS) mass for $b$-quark, since it appears in the heavy quark expansion and the  PS mass is less ambiguous than the pole mass.
\begin{table}[htb]
\begin{center}
\caption{Imput parameters.}
\begin{tabular}{lccc}
\hline
\hline	
        $m_b^{\rm PS}$ & 4.53 {\rm Gev} & $V_{cb}$ & 4.1$\times10^{-2}$\\
        $m_{\Lambda_b}$ & 5.620 {\rm GeV} & $m_{\Lambda_c}$ & 2.286 {\rm GeV}\\
		$s_0$ & $10\pm 0.5$ {\rm GeV}$^2$& $M^2$ & $7.5\pm 2.5$ {\rm GeV}$^2$ \\
		$f^{(1)}_{\Lambda_b}$& $0.030\pm 0.005$ {\rm GeV}$^3$& $f^{(2)}_{\Lambda_b}$ & $0.03 \pm 0.003$  {\rm GeV}$^3$\\
		$\lambda^{A}_{\Lambda_c}$ & $(1.51\pm0.35)\times 10^{-2}$ {\rm GeV}$^2$ & $\lambda^{P}_{\Lambda_c}$ & $(1.19\pm0.19)\times 10^{-2}$ {\rm GeV}$^2$\\
\hline
\hline
\end{tabular}		
\end{center}\label{Tab:input}
\end{table}

Since the LCSR is valid only at small $q^2$, we first present the results of the form factors $f_1$ and $f_2$ at $q^2=0$, which are displayed in Table \ref{Tab:F0}. In order to highlight the power suppressed contribution from the heavy quark expansion, both the leading power contribution and NLP contribution are listed for a comparison, and it is obvious the NLP contribution can reduce the leading power contribution about 20 \%, which will significantly change the results of the physical observables. Of cause we should note that the power corrections considered in this paper is very preliminary, it is  necessary to perform a more careful treatment of the NLP contributions.  In this table, the form factors $f_1(0)$ and $f_2(0)$ are evaluated with both P-type and A-type interpolation currents, and the results indicate that A-type current leads to a larger results for all the three models of the LCDAs of $\Lambda_b$-baryon, and in general they can be consistent within the error area. The total uncertainties shown in this table are obtained by varying separate input parameters within their ranges and adding the resulting separate uncertainties of the form factors in quadrature. The results from QCDSR model and free parton model of $\Lambda_b$ LCDAs are  well consistent with each other, and results from the exponential model are smaller for both A-type and P-type currents. The result of the form factor $f_3(0)$ still satisfies $f_3(0)=0$ which is expected in heavy $b$-quark limit. Although we have considered the power correction from heavy quark expansion, it does not yield nonzero contribution to $f_3(0)$. The relations between the form factors displayed in Eq.(\ref{ffrelation}) which are from the heavy quark symmetry are still valid as shown the numerical results in Table \ref{Tab:F0}. 
%*****************************************************
\begin{table}[htb]
  \centering
    \caption{Form factors $f^{P}_i(0)$ and $f^{A}_i(0)$ at $q^2=0$;}
		\begin{tabular}{c c c c c c c}
		\hline
		\hline	
		{\rm Model} & $f^{\rm LP}_1$&$f^{\rm NLP}_1$ &$f^{\rm LP+NLP}_1$& $f^{\rm LP}_2$& $f^{\rm NLP}_2$&$f^{\rm LP+NLP}_2$ \\
		\hline 
		\textbf{A-type current}\\
		{\rm QCDSR}&$0.977$&$-0.188$&$0.789\pm 0.243$&$-0.303$&$0.059$&$-0.244\pm 0.076$ \\
		{\rm Exponential}&$0.859$&$-0.162$&$0.697\pm0.297$&$-0.265$&$0.050$&$-0.215\pm0.094$\\
		{\rm Free parton}&$0.931$&$-0.182$&$0.749\pm0.339$&$-0.287$&$0.057$&$-0.230\pm0.108$ \\
	    \hline
	    \textbf{P-type current}\\
	    {\rm QCDSR}&$0.854$&$-0.168$&$0.686\pm0.169$&$-0.289$&$0.060$&$-0.229\pm0.059$\\
	    {\rm Exponential}&$0.711$&$-0.137$&$0.574\pm0.214$&$-0.231$&$0.047$&$-0.184\pm0.061$\\
		{\rm Free parton}&$0.851$&$-0.171$&$0.680\pm0.248$&$-0.300$&$0.063$&$-0.237\pm0.079$ \\
		 \hline
		 & $g^{\rm LP}_1$&$g^{\rm NLP}_1$ &$g^{\rm LP+NLP}_1$& $g^{\rm LP}_2$& $g^{\rm NLP}_2$&$g^{\rm LP+NLP}_2$ \\
		 \hline
		 \textbf{A-type current}\\
		{\rm QCDSR}&$0.675$&$-0.130$&$0.545\pm0.167$&$-0.303$&$0.059$&$-0.244\pm0.076$\\
		 {\rm Exponential}&$0.594$&$-0.112$&$0.482\pm0.203$&$-0.265$&$0.050$&$-0.215\pm0.094$\\
		 {\rm Free parton}&$0.644$&$-0.125$&$0.519\pm0.231$&$-0.287$&$0.057$&$-0.230\pm0.108$\\
		 \hline
		 \textbf{P-type current}\\
		 {\rm QCDSR}&$0.565$&$-0.108$&$0.457\pm0.111$&$-0.289$&$0.060$&$-0.229\pm0.059$\\
		 {\rm Exponential}&$0.480$&$-0.090$&$0.390\pm0.155$&$-0.231$&$0.047$&$-0.184\pm0.061$\\
		 {\rm Free parton}&$0.551$&$-0.108$&$0.443\pm0.171$&$-0.300$&$0.063$&$-0.237\pm0.079$\\
		\hline
		\hline
		\end{tabular}
\label{Tab:F0}
\end{table}
%*****************************************************
%==================================
\begin{table}[htb]
	\begin{center}
		\caption{The results for form factor at $q^2=0$ of this work are compared with those of other methods}
		\begin{tabular}{ccccccc}
			\hline & $f_1(0)$& $f_2(0)$&$f_3(0)$& $g_1(0)$&$g_2(0)$&$g_3(0)$ \\
			\hline 	
				{\rm This work(A-type )}&&&&&&\\
		{\rm QCDSR model}	&$0.789$&$-0.244$&$0$&$0.545$&$-0.244$&$0$\\
			{\rm Exponential  model}&$0.697$ & $-0.215$&$0$& $0.482$&$-0.215$&$0$\\
				{\rm Free parton model }	&$0.749$&$-0.230$&$0$&$0.519$&$-0.230$&$0$\\
			\hline 	
				{\rm This work(P-type )}&&&&&&\\
		{\rm QCDSR model}	&$0.686$&$-0.229$&$0$&$0.457$&$-0.229$&$0$\\
			{\rm Exponential  model}&$0.574$ & $-0.184$&$0$& $0.390$&$-0.184$&$0$\\
				{\rm Free parton model }	&$0.680$&$-0.237$&$0$&$0.443$&$-0.237$&$0$\\
				\hline
			{\rm QCDSR\cite{Zhao:2020mod}}&$0.604$& $-0.101$&$-0.059$& $0.456$&$-0.124$&$0.080$\\
			{\rm LFQM\cite{Zhu:2018jet}}&$0.638$& $-0.107$&$-0.036$& $0.500$&$-0.100$&$0.028$  \\
			{\rm LFQM\cite{Li:2021qod}}&$0.669$ & $-0.160$&$-0.033$& $0.478$&$-0.170$&$0.053$ \\
			{\rm RQM\cite{Faustov:2016pal}}&$0.719$ & $-0.212$&$-0.025$& $0.521$&$-0.279$&$0.091$ \\
			{\rm CCQM\cite{Gutsche:2015mxa}}&$0.704$ & $-0.133$&$-0.035$& $0.531$&$-0.141$&$0.042$ \\
			{\rm LQCD\cite{Detmold:2015aaa}}&$0.558$& $-0.174$&$-0.010$& $0.388$&$-0.210$&$0.082$  \\
			\hline
		\end{tabular}
		\label{ffcomparison}
	\end{center}
\end{table}
%=====================================
In Table \ref{ffcomparison}, we collected the predictions of the form factor $f_i$ at $q^2=0$ from the light-front quark model\cite{Zhu:2018jet,Li:2021qod}, the relativistic quark model\cite{Faustov:2016pal}, the covariant constituent quark model\cite{Gutsche:2015mxa}, the QCD sum rule\cite{Zhao:2020mod} and the Lattice QCD simulation\cite{Detmold:2015aaa}, together with our results.  Lattice simulation is valid at large $q^2$, the prediction here depends on the extrapolation model and it is  smaller than the other predictions, which leads to  too small branching ratios of the semileptonic decays compared with the experimental measurement. The different predictions is in general consistent with each other if the uncertainties are taken into account, and in our calculation there are two preferable scenarios: the A-type interpolation current  together with the exponential model of the DAs of $\Lambda_b$ and the P-type interpolation current together with the QCDSR model or free parton model of the DAs of $\Lambda_b$ .  Therefore it is hard to distinguish different models or the interpolation current from the predictions of the form factors from the current calculation. The form factors $g_i$ is directly related to $f_i$, so we will not give more discussions.

\begin{figure}%[h]
\begin{center}
\includegraphics[width=0.47 \columnwidth]{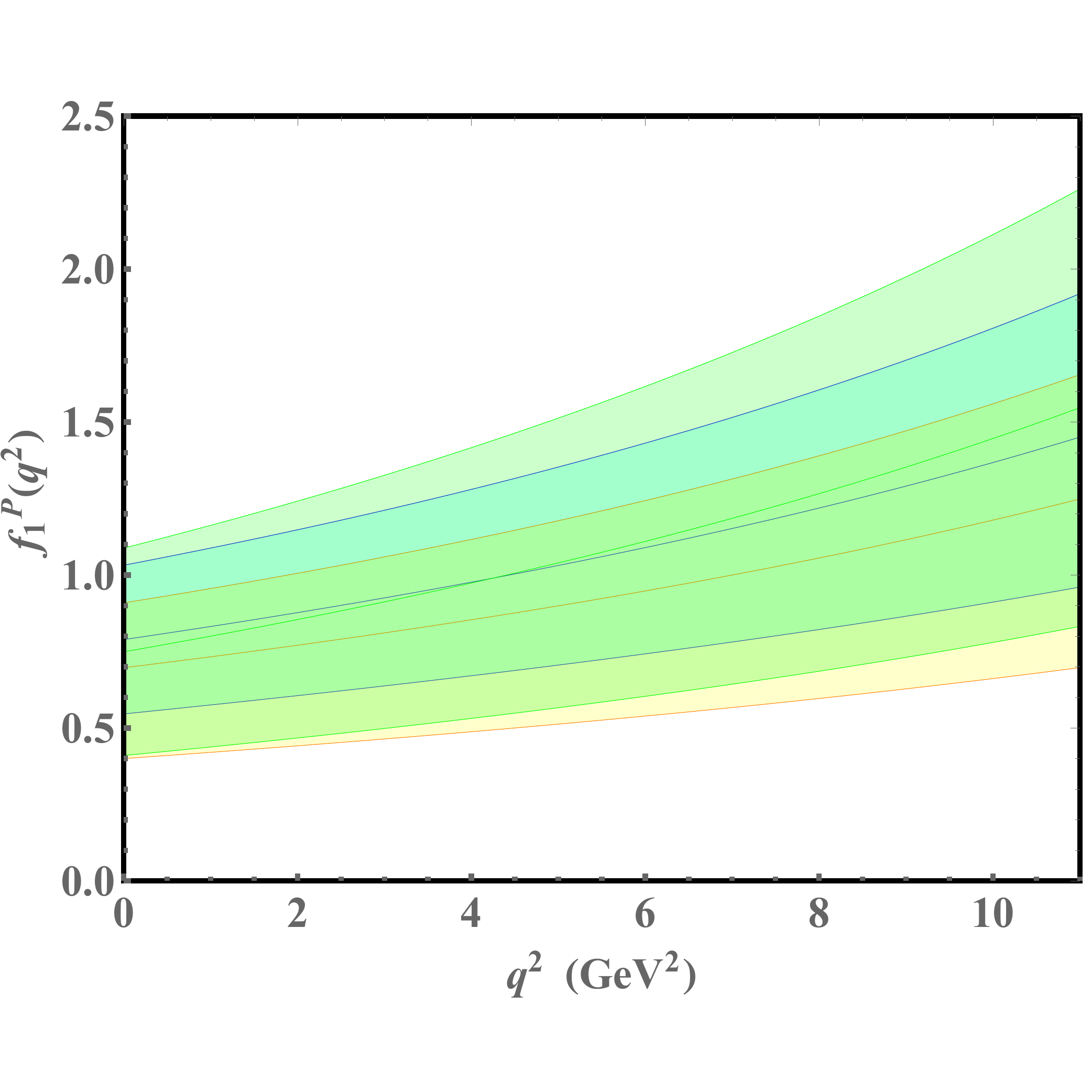}
\includegraphics[width=0.47 \columnwidth]{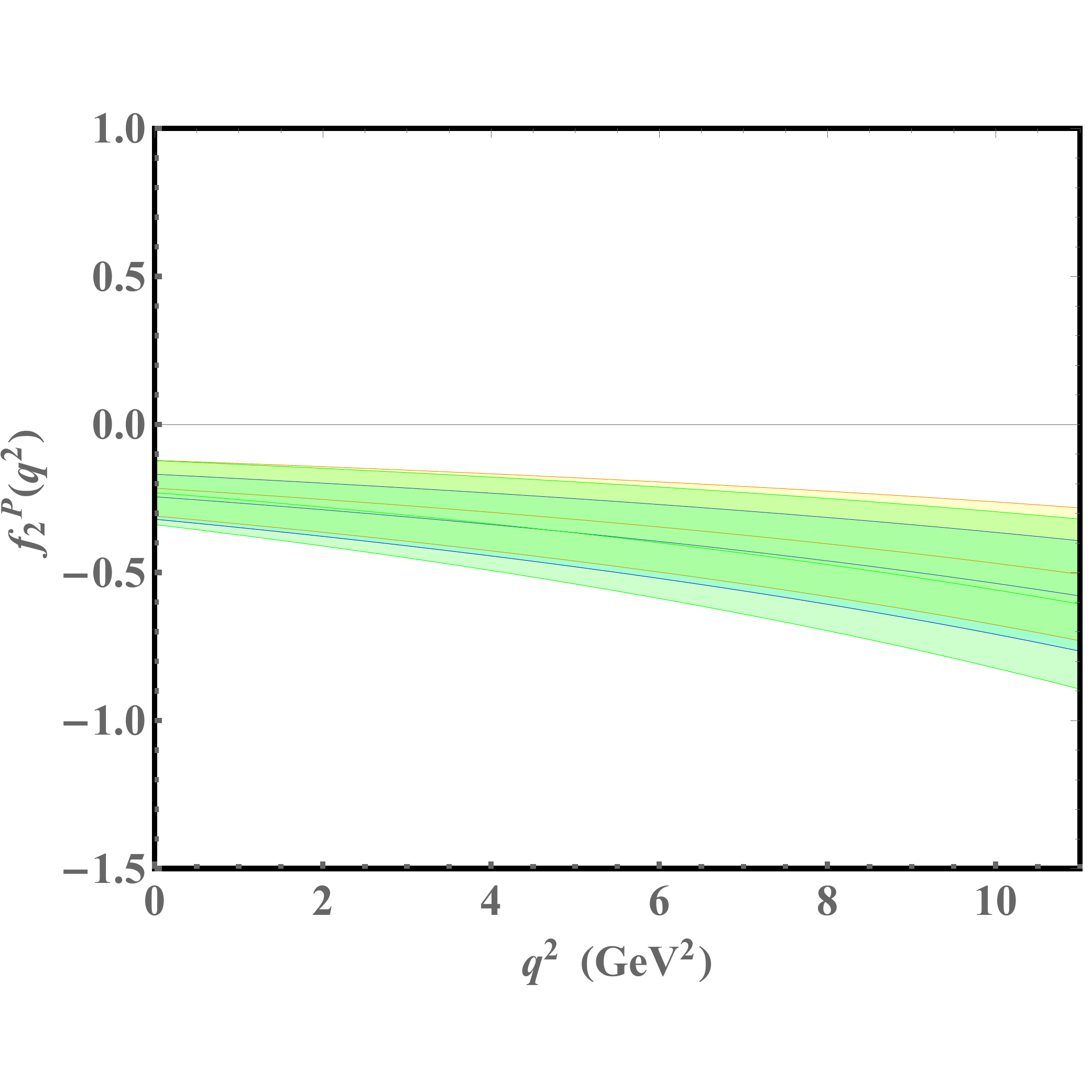}
\vspace*{0.1cm}
\caption{The form factors $f_1(q^2)$ and $f_2(q^2)$ from A-type interpolation current. The blue band, yellow band and green band denote the the form factors with the QCDSR model, the exponential model and the free parton model adopted respectively.}
\label{fig: q2depend}
\end{center}
\end{figure}

In order to predict the experimental observables, we extrapolate our results in small $q^2$($0\leq 5{\rm GeV^2}$) to the whole physical region.   To this end, we employ the simplified $z$-series parametrization
\cite{Bourrely:2008za} based upon the conformal mapping
\begin{eqnarray}
z(q^2, t_0) = \frac{\sqrt{t_{+}-q^2}-\sqrt{t_{+}-t_0}}{\sqrt{t_{+}-q^2}+\sqrt{t_{+}-t_0}},\,
\end{eqnarray}
which transforms the  cut $q^2$-plane onto the disk $|z(q^2, t_0)| \leq 1$ in the complex $z$-plane.
We choose the parameter $t_{\pm}$ to be $t_{\pm}=(m_{\Lambda_b}\pm m_{\Lambda_c})^2$, and $t_0=t_+-\sqrt{t_+-t_{-}}\sqrt{t_+-t_{min}}$ in order to reduce the interval of $z$ after mapping $q^2$ to $z$ with the interval $t_{min}<q^2<t_{-}$. In the numerical analysis, we take $t_{min}= -6$ GeV$^2$.
Keeping the series expansion of the form factors to the first power of $z$-parameter,
we propose the following parameterizations
\begin{eqnarray}
f_i(q^2)&=& \frac{f_i(0)}{
1-q^2/m_{B_c^{\ast}(1^-)}^2} \,
\left \{ 1 + a_1^{i} \, \left [ z(q^2, t_0) - z(0, t_0) \right ] \right \}  \,\nonumber \\
g_i(q^2)&=& \frac{g_i(0)}{1-q^2/m_{B_c^{\ast}(1^+)}^2} \,
\left \{ 1 + b_1^{i} \, \left [ z(q^2, t_0) - z(0, t_0) \right ] \right \},  \,
\end{eqnarray}
where the mass of ${B_c^{\ast}(1^-)}$ and ${B_c^{\ast}(1^+)}$ appears in the pole factor, while they have not been measured. There are some theoretical estimations on these masses, and here we adopt $m_{B_c^{\ast}(1^-)}=6.336{\rm GeV}$, $m_{B_c^{\ast}(1^+)}=6.745{\rm GeV}$ \cite{Bernlochner:2018bfn}. The Fitted results of $ a_1^i, b_1^i$  are given  Table \ref{zpara}. Since the  form factors have been extrapolated to the whole physical region, we plot the $q^2$-dependence of the form factors  with different DAs of $\Lambda_b$ baryon in Fig. \ref{fig: q2depend}. The  uncertainties shown in the bands are obtained by  adding the resulting separate uncertainties from $f_i(0),a_i,b_i$ in quadrature.

%%%%%%%%%%%
%%%%%%%%%%%

In the following, we aim at exploring phenomenological applications of the obtained $\Lambda_b \to \Lambda_c$ form factors which serve as fundamental ingredients for the theory description of the $\Lambda_b \to \Lambda_c \ell \nu $ decays which are regarded as a good  platform to further investigate the $R_{D(D^\ast)}$ anomaly. In order to calculate the phenomenological observables  such as the branching ratios, the forward backward asymmetries, etc., it is convenient to introduce the helicity amplitudes which are defined by
\begin{eqnarray}
	H^{V,A}_{\lambda_{\Lambda_c},\lambda_{W^-} }=\epsilon^{\dagger\mu}(\lambda_{W^-})\langle \Lambda_c,\lambda_{\Lambda_c}|V(A)|\Lambda_b, \lambda_{\Lambda_b}\rangle ,
\end{eqnarray}
where the $\lambda_{\Lambda_b}, \lambda_{\Lambda_c}, \lambda_{W^-}$ denote the helicity of the $\Lambda_b$ baryon, the $\Lambda_c$ baryon and the off-shell $W^-$ which mediates the semileptonic decays, respectively. The helicity amplitudes $H^{V,A}_{\lambda_{\Lambda_c},\lambda_{W^-}}$ can be expressed as functions of the form factors
\begin{eqnarray}
	H^{V}_{\frac{1}{2},0}&=&\frac{\sqrt{Q_{-}}}{\sqrt{q^2}}\Big[M_{+}F_1(q^2)- \frac{q^2}{m_{\Lambda_b}}F_2(q^2)\Big],\,\,\,
	H^{A}_{\frac{1}{2},0}=\frac{\sqrt{Q_{+}}}{\sqrt{q^2}}\Big[M_{-}G_1(q^2)+ \frac{q^2}{m_{\Lambda_b}}G_2(q^2)\Big] , \nonumber \\
	H^{V}_{\frac{1}{2},1}&=&\sqrt{2Q_{-}}\Big[F_1(q^2)- \frac{M_{+}}{m_{\Lambda_b}}F_2(q^2)\Big] , \,\,\,
	H^{A}_{\frac{1}{2},1}=\sqrt{2Q_{+}}\Big[G_1(q^2)+ \frac{M_{-}}{m_{\Lambda_b}}G_2(q^2)\Big] , \nonumber \\
	H^{V}_{\frac{1}{2},t}&=&\frac{\sqrt{Q_{+}}}{\sqrt{q^2}}\Big[M_{-}F_1(q^2)+ \frac{q^2}{m_{\Lambda_b}}F_3(q^2)\Big], \,\,\,
	H^{A}_{\frac{1}{2},t}=\frac{\sqrt{Q_{-}}}{\sqrt{q^2}}\Big[M_{+}G_1(q^2)- \frac{q^2}{m_{\Lambda_b}}G_3(q^2)\Big] .
\end{eqnarray}
where $Q_{\pm}$ is defined as $Q_{\pm}=(m_{\Lambda_b}\pm m_{\Lambda_c})^2-q^2$ and $M_{\pm}=m_{\Lambda_b} \pm m_{\Lambda_c}$. The negative helicities can be obtained by
\begin{eqnarray}
    H^{V}_{-\lambda_{\Lambda_c},-\lambda_{W^-}}=H^{V}_{\lambda_{\Lambda_c},\lambda_{W^-}}\,\ \ \  H^{A}_{-\lambda_{\Lambda_c},-\lambda_{W^-}}=-H^{A}_{\lambda_{\Lambda_c},\lambda_{W^-}} .
\end{eqnarray}
The total helicity amplitudes are then written by
\begin{eqnarray}
    H_{\lambda_{\Lambda_c},\lambda_{W^-}}=H^{V}_{\lambda_{\Lambda_c},\lambda_{W^-}}-H^{A}_{\lambda_{\Lambda_c},\lambda_{W^-}},
\end{eqnarray}
The differential angular distribution for the decay $\Lambda_b
\rightarrow \Lambda_c\ell \bar{\nu}_\ell$ has the following form
\begin{eqnarray}
	\frac{d\Gamma(\Lambda_b \rightarrow \Lambda_c \ell^-\bar{\nu}_\ell)}{dq^2d \cos\theta_\ell}=\frac{G_F^2|V_{cb}|^2q^2|\vec{p^\prime}|}{512\pi^3m^2_{\Lambda_b}}\Big(1-\frac{m^2_\ell}{q^2}\Big)^2
\Big(A_1+\frac{m_\ell^2}{q^2}A_2\Big),
\end{eqnarray}
where $G_F$ is the Fermi constant, $V_{cb}$ is the CKM matrix element, $m_\ell$ is the lepton mass($\ell=e,\mu,\tau$), $\theta_\ell$ is the angle between the three-momentum  of the final $\Lambda_c$ baryon and the lepton in the $q^2$ rest frame, $\vec{p^\prime}$ is the three-momentum of $\Lambda_c$ baryon, and the amplitudes $A_i$ are defined as
\begin{eqnarray}
	A_1&=&2\sin^2\theta_\ell(H^2_{1/2,0}+H^2_{-1/2,0})+(1-\cos \theta_\ell)^2H^2_{1/2,1}+(1+\cos\theta_\ell)^2H^2_{-1/2,-1} ,\nonumber \\
	A_2&=&2\cos^2\theta_\ell(H^2_{1/2,0}+H^2_{-1/2,0})+\sin^2 \theta_\ell(H^2_{1/2,1}+H^2_{-1/2,-1})+2(H^2_{1/2,t}+H^2_{-1/2,t}) \nonumber \\
	&\ &-4\cos\theta_\ell(H_{1/2,t}H_{1/2,0}+H_{-1/2,t}H_{-1/2,0}), 	%|\vec{p^\prime}|&=&\frac{\sqrt{m_{\Lambda_b}^4+m^4_{\Lambda_c)}+q^2-2(m^2_{\Lambda_b}m^2_{\Lambda_c}+m^2_{\Lambda_c}q^2+m^2_{\Lambda_b}q^2)}}{2m_{\Lambda_b}} ,
\end{eqnarray}
The differential decay rate can be obtained by integrating out $\cos\theta_l$
\begin{eqnarray}
	\frac{d\Gamma(\Lambda_b\rightarrow \Lambda_c\ell^-\bar{\nu}_l)}{dq^2}=\int^1_{-1}\frac{d\Gamma(\Lambda_b \rightarrow \Lambda_c\ell^-\bar{\nu}_\ell)}{dq^2d\cos\theta_\ell}d\cos\theta_\ell.
\end{eqnarray}
In addition, the  other observables such as leptonic forward-backward asymmetry ($A_{FB}$), the final state hadron polarization($P_B$) and the lepton polarization($P_\ell$), are defined as
\begin{equation}
\begin{aligned}	
	A_{FB}(q^2)&=\frac{\int^1_0 \frac{d\Gamma}{dq^2d\cos\theta_\ell}d\cos\theta_l-\int^0_{-1} \frac{d\Gamma}{dq^2d\cos\theta_\ell}d\cos\theta_\ell}{\int^1_0 \frac{d\Gamma}{dq^2d\cos\theta_\ell}d\cos\theta_\ell+\int^0_{-1} \frac{d\Gamma}{dq^2d\cos\theta_\ell}d\cos\theta_\ell} , \\	P_B(q^2)&=\frac{d\Gamma^{\lambda_{\Lambda_c}=1/2}/dq^2-d\Gamma^{\lambda_{\Lambda_c}=-1/2}/dq^2}{d\Gamma/dq^2} , \\
	P_\ell(q^2)&=\frac{d\Gamma^{\lambda_\ell=1/2}/dq^2-d\Gamma^{\lambda_\ell=-1/2}/dq^2}{d\Gamma/dq^2},
\end{aligned}
\end{equation}
 and the differential widths with definite polarization of the final state can be written by
\begin{equation}
\begin{aligned}	\frac{d\Gamma^{\lambda_{\Lambda_c}=1/2}}{dq^2}&=\frac{4m^2_l}{3q^2}\Big(H^2_{1/2,1}+H^2_{1/2,0}+3H^2_{1/2,t}\Big)+\frac{8}{3}\Big(H^2_{1/2,0}+H^2_{1/2,1}\Big) , \\
	\frac{d\Gamma^{\lambda_{\Lambda_c}=-1/2}}{dq^2}&=\frac{4m^2_l}{3q^2}\Big(H^2_{-1/2,-1}+H^2_{-1/2,0}+3H^2_{-1/2,t}\Big)+\frac{8}{3}\Big(H^2_{-1/2,0}+H^2_{-1/2,-1}\Big) , \\		\frac{d\Gamma^{\lambda_\ell=1/2}}{dq^2}&=\frac{m^2_l}{q^2}\Big[\frac{4}{3}\Big(H^2_{1/2,1}+H^2_{1/2,0}+H^2_{-1/2,-1}+H^2_{-1/2,0}\Big)+4\Big(H^2_{1/2,t}+H^2_{-1/2,t}\Big)\Big] , \\
	\frac{d\Gamma^{\lambda_\ell=-1/2}}{dq^2}&=\frac{8}{3}\Big(H^2_{1/2,1}+H^2_{1/2,0}+H^2_{-1/2,-1}+H^2_{-1/2,0}\Big).
\end{aligned}
\end{equation}

 The numerical results of the relevant observables in the semi-leptonic decays  $\Lambda_b \rightarrow \Lambda_c\ell\nu$ are presented  in Table. \ref{observable} where both the A-type and P-type interpolation current are considered. Three different models of $\Lambda_b$ baryon are employed in the calculation so that they can be compared with the experimental result to determine which one is more preferable. The central value of the life time of $\Lambda_b$  is adopted as $\tau_{\Lambda_b}=1.470$ps and   the CKM matrix element $|V_{cb}|$ has been present in Table \ref{Tab:input}. From the Table. \ref{observable}, we can see that the integrated branching ratio for the semi-leptonic decay  $\Lambda_b \rightarrow \Lambda_c \ell^-\nu$ from P-type interpolating currents is slight smaller than that from A-type current. Compared with the experimental data $Br(\Lambda_b \rightarrow \Lambda_c\ell\nu)=6.2^{+1.4}_{-1.3}\%$, the prediction of A-type operators seems more consistent with the data if the exponential  model is adopted.  We note that our result is from the tree level calculation of the leading power contribution in the heavy quark limit plus a rough estimation of the power corrections from the heavy quark expansion, this conclusion is very preliminary and a more careful study is required to distinguish different models of $\Lambda_b$ DAs and the interpolation currents.  The numerical results of leptonic forward-backward asymmetry ($A_{FB}$), the final hadron polarization($P_B$) and the lepton polarization($P_\ell$) are also  presented in the Table \ref{observable}. Since these observables are not very sensitive to the form factors at small $q^2$ region, the predictions from different models of LCDAs and different interpolation currents are very close.  To compare our results and the prediction from the other methods, we collect the numerical results from various studies in Table \ref{comparison of observables}. We can see that the integrated branching ratio from various studies does not significantly deviate form each other, while the other observables are more sensitive to different approaches, which can serve as the basis to distinguish different methods.  We also present the ratio of branching ratio $R_{\Lambda_c}$ in the Table \ref{observable}, it is not very sensitive to the interpolation current and the model of the LCDAs of $\Lambda_b$, and the central value of our prediction is a little smaller than some recent study\cite{Bernlochner:2018kxh}, but can be consistent with the recent  LHCb reported result $R(\Lambda_c) =0.242 \pm 0.026 \pm 0.040 \pm 0.059$ \cite{LHCb:2022piu}. Our predictions for the branching ratios have large uncertainty, to improve the theoretical precision, one can make progress in the following two aspects: one is to reduce the uncertainty of the parameters inside the DAs of heavy baryon by global fit or the Lattice calculation, the other is to include the loop corrections and more power corrections. 

%======================================================
\begin{table}[htb]
  \centering
   \caption{The fitted results of $a_1,b_1 $ for the form factors $F^{p}_i$ ,$F^{A}_i$ , $G^{p}_i$ and  $G^{p}_i$}
		\begin{tabular}{c c c c c c}
		\hline
		\hline
		{\rm Model} && $f_1$& $f_2$&$g_1$&$g_2$  \\
		\hline
		\multicolumn{6}{c}{\textbf{A-type current}}\\
		\hline
		\multirow{2}{*}{\rm QCDSR}&$f(0)$&$0.789\pm0.243$&$-0.244\pm0.076$&$0.545\pm0.167$&$-0.244\pm0.076$ \\
		&$a_1$&$-6.92\pm2.7$&$-14.95\pm2.9$&$-4.15\pm2.8$&$-15.93\pm2.9$\\
		\hline
		\multirow{2}{*}{\rm Exponential}&$f(0)$&$0.697\pm0.297$&$-0.215\pm0.094$&$0.482\pm0.203$&$-0.215\pm0.094$\\
		&$a_1$&$-6.21\pm3.1$&$-14.65\pm2.6$&$-3.26\pm2.6$&$-15.63\pm2.5$\\
		\hline
		\multirow{2}{*}{\rm Free parton}&$f(0)$&$0.749\pm0.339$&$-0.230\pm0.108$&$0.519\pm0.231$&$-0.230\pm0.108$\\
		&$a_1$&$-10.32\pm2.9$&$-18.91\pm2.4$&$-7.37\pm3.0$&$-19.94\pm2.3$\\
		\hline
		\multicolumn{6}{c}{\textbf{P-type current}}\\
			\hline
		\multirow{2}{*}{\rm QCDSR}&$f(0)$&$0.686\pm0.169$&$-0.229\pm0.059$&$0.457\pm0.111$&$-0.229\pm0.059$ \\
		&$b_1$&$-5.74\pm2.2$&$-11.00\pm2.4$&$-3.96\pm2.1$&$-11.93\pm2.4$\\
		\hline
		\multirow{2}{*}{\rm Exponential}&$f(0)$&$0.574\pm0.214$&$-0.184\pm0.061$&$0.390\pm0.155$&$-0.184\pm0.061$\\
		&$b_1$&$-6.25\pm3.1$&$-12.68\pm2.6$&$-4.04\pm2.8$&$-13.62\pm2.6$\\
		\hline
		\multirow{2}{*}{\rm Free parton}&$f(0)$&$0.680\pm0.248$&$-0.237\pm0.079$&$0.443\pm0.171$&$-0.237\pm0.079$\\
		&$b_1$&$-6.92\pm2.6$&$-12.21\pm2.8$&$-4.96\pm2.8$&$-13.16\pm2.7$\\
		\hline
		\hline
		\end{tabular}\label{zpara}
\end{table}
%=======================================================
\begin{table}[h]
	\centering
	\caption{The predictions for the branching fractions, the averaged leptonic forward-backward asymmetry$\langle A_{FB}\rangle$, the averaged final hadron polarization $\langle P_{B}\rangle$ and the averaged lepton polarization $\langle P_{l}\rangle$ for $\Lambda_b\rightarrow \Lambda_cl^-\bar{\nu}_l$ under two interpolating current(A-type and P-type) with three different LCDA models of $\Lambda_b$ baryon(QCDSR, Exponential and Free-parton).}
	\begin{tabular}{c c c c c c c}
		\hline
		\hline
		Model&$l$&Br($\times 10^{-2}$)&$\langle A_{FB}\rangle$&$\langle P_B\rangle$&$\langle P_l\rangle$& $R_{\Lambda_c}$\\
		\hline
		\multicolumn{7}{c}{\textbf{A-type current}}\\
		\hline
		\multirow{3}{*}{QCDSR}&e&$7.68\pm3.66$&$0.18\pm0.02$&$-0.87\pm0.15$&$-1.00\pm0.00$&\multirow{3}{*}{$0.273\pm0.013$}\\
		&$\mu$&$7.65\pm3.65$&$0.17\pm0.02$&$-0.87\pm0.15$&$-0.98\pm0.00$\\
		&$\tau$&$2.09\pm1.05$&$-0.05\pm0.03$&$-0.79\pm0.14$&$-0.29\pm0.17$\\
		\hline
		\multirow{3}{*}{Exponential}&e&$5.81\pm3.78$&$0.18\pm0.03$&$-0.88\pm0.20$&$-1.00\pm0.00$& \multirow{3}{*}{$0.268\pm0.015$}\\
		&$\mu$&$5.78\pm3.77$&$0.17\pm0.03$&$-0.87\pm0.20$&$-0.98\pm0.01$\\
		&$\tau$&$1.55\pm1.06$&$-0.05\pm0.05$&$-0.79\pm0.19$&$-0.29\pm0.24$ \\
		\hline
		\multirow{3}{*}{Free-parton}&e&$7.85\pm5.43$&$0.18\pm0.03$&$-0.86\pm0.22$&$-1.00\pm0.00$& \multirow{3}{*}{$0.288\pm0.016$}\\
		&$\mu$&$7.82\pm5.41$&$0.18\pm0.03$&$-0.86\pm0.22$&$-0.98\pm0.01$\\
		&$\tau$&$2.25\pm1.62$&$-0.04\pm0.05$&$-0.78\pm0.21$&$-0.30\pm0.25$\\
		\hline
		\multicolumn{7}{c}{\textbf{P-type current}}\\
		\hline
		\multirow{3}{*}{QCDSR}&e&$5.42\pm2.06$&$0.18\pm0.02$&$-0.88\pm0.12$&$-1.00\pm0.00$& \multirow{3}{*}{$0.270\pm0.011$}\\
		&$\mu$&$5.40\pm2.05$&$0.17\pm0.02$&$-0.87\pm0.12$&$-0.98\pm0.00$\\
		&$\tau$&$1.46\pm0.58$&$-0.05\pm0.03$&$-0.79\pm0.11$&$-0.29\pm0.14$\\
		\hline
		\multirow{3}{*}{Exponential}&e&$3.93\pm2.37$&$0.18\pm0.03$&$-0.87\pm0.18$&$-1.00\pm0.00$& \multirow{3}{*}{$0.271\pm0.014$}\\
		&$\mu$&$3.91\pm2.36$&$0.17\pm0.03$&$-0.87\pm0.18$&$-0.98\pm0.01$\\
		&$\tau$&$1.06\pm0.67$&$-0.05\pm0.04$&$-0.79\pm0.17$&$-0.29\pm0.21$\\
		\hline
		\multirow{3}{*}{Free-parton}&e&$5.36\pm3.15$&$0.19\pm0.03$&$-0.88\pm0.18$&$-1.00\pm0.00$& \multirow{3}{*}{$0.274\pm0.014$}\\
		&$\mu$&$5.34\pm3.14$&$0.18\pm0.03$&$-0.87\pm0.18$&$-0.98\pm0.01$\\
		&$\tau$&$1.47\pm0.90$&$-0.05\pm0.04$&$-0.80\pm0.17$&$-0.29\pm0.21$\\
		\hline
		\hline
	\end{tabular}
	\label{observable}
\end{table}
%======================================================
\begin{table}[htb]
	\centering
	\caption{The predictions for the branching fractions, the averaged leptonic forward-backward asymmetry$\langle A_{FB}\rangle$, the averaged final hadron polarization $\langle P_{B}\rangle$ and the averaged lepton polarization $\langle P_{l}\rangle$ for $\Lambda_b\rightarrow \Lambda_cl^-\bar{\nu}_l$ under different methods.}
	\begin{tabular}{c c c c c c}
		\hline
		\hline
		&$l$&Br($\times 10^{-2}$)&$\langle A_{FB}\rangle$&$\langle P_B\rangle$&$\langle P_l\rangle$ \\
		\hline
		\multirow{3}{*}{$\begin{aligned}&{\rm This\,\,\, work} \\ &{\rm (A-type\,\,\, current,Exponential\,\,\, model)}\end{aligned}$}&$e$&$5.81$&$0.18$&$-0.88$&$-1.00$ \\
		&$\mu$&$5.78$&$0.17$&$-0.87$&$-0.98$ \\
		&$\tau$&$1.55$&$-0.05$&$-0.79$&$-0.29$ \\
		\hline
		\multirow{3}{*}{RQM\cite{Faustov:2016pal}}&e&$6.48$&$0.195$&-&-\\
		&$\mu$&$6.46$&$0.189$&-&-\\
		&$\tau$&$2.03$&$-0.021$&-&-\\
		\hline
		\multirow{3}{*}{LFQM\cite{Li:2021qod}}&e&-&$0.18$&$-0.81$&$-1.00$\\
		&$\mu$&-&$0.17$&$-0.81$&$-0.98$\\
		&$\tau$&-&$-0.08$&$-0.77$&$-0.24$\\
		\hline
		\multirow{3}{*}{CCQM\cite{Gutsche:2015mxa}}&e&$6.9$&$0.36$&-&-\\
		&$\mu$&-&-&-&-\\
		&$\tau$&$2.0$&$-0.077$&-&-\\
		\hline
		\hline
	\end{tabular}
	\label{comparison of observables}
\end{table}
%======================================================

\section{Summary}
We have calculated the form factors of $\Lambda_b \rightarrow \Lambda_c$ transition within the framework of LCSR with the DAs of $\Lambda_b$-baryon, and further investigated  the experimental observables such as the branching ratios, the forward-backward asymmetries, the final state polarizations of the semileptonic decays $\Lambda_b \rightarrow \Lambda_c\ell \nu$ and the ratio of the branching ratios $R_{\Lambda_c}$. Since the interpolating current of the baryon is not unique, we employed P-type and A-type interpolation current for a cross check of our predictions. Following a standard procedure of the calculation of heavy-to-light form factors by using LCSR approach, we can arrive at the sum rules of the $\Lambda_b \rightarrow \Lambda_c$ transition form factors. In the hadronic representation of the correlation function, we have included $\Lambda_c^*$ state  in addition to $\Lambda_c$ state so that the  $\Lambda_b \rightarrow \Lambda_c$ form factors can be evaluated without ambiguity. The LCDAs of $\Lambda_b$-baryon are not well determined so far, thus we employed three different models, i.e, the QCDSR model, the exponential model, the free-parton model for a comparison.

Since the DAs of the $\Lambda_b$ baryon are defined in term of the large component of $b$-quark field in HQET, a direct calculation will lead to the form factors at heavy $b$-quark limit, and only two of them are independent. To improve the accuracy of the predictions, we include the power suppressed contribution from the power suppressed bottom quark field in the heavy quark expansion. However, we neglected the contribution from four-particle DAs of the $\Lambda_b$ baryon since there is no studies on these DAs so far. As as result, the power suppressed contribution considered in this paper does not change the form factor relations in the heavy $b$ quark limit. Numerically, the power suppressed contribution reduced the leading power result about $20\%$. The total results of the form factors from the P-type interpolation current is smaller than that from the A-type interpolation current, it is hard to distinguish which one is more preferable since the result also depends on the DAs of the $\Lambda_b$ baryon. The LCSR is valid at small $q^2$ region, thus we extrapolate our results to the whole physical region using $z$-series expansion, then we can obtain the $q^2$ dependence of the form factors which is important to predict the experimental observables. 

We further obtained the predictions of the total branching fractions, the averaged forward-backward asymmetry $\langle A_{FB}\rangle$, the averaged final hadron polarization $\langle P_{B}\rangle$ and the averaged lepton polarization $\langle P_{l}\rangle$ of the $\Lambda_b \to \Lambda_c \ell\mu$ decays, as well as the ratio of branching ratios $R_{\Lambda_c}$.  Our predicted branching ratios from the A-type interpolation current are more close to the experimental data once the exponential model of the DAs of $\Lambda_b$-baryon is adopted, and  they are also consistent with the predictions from the relativistic quark model, the light-front quark model, etc.. The ratio of branching ratio $R_{\Lambda_c}$  is not very sensitive to the interpolation current and the model of the LCDAs of $\Lambda_b$, and the central value of our prediction can be  consistent with the recent data of LHCb.  Moreover, we only performed a tree-level calculation of the correlation function, and the QCD corrections to the hard kernel in the partonic expression of the correlation function are needed to increase the accuracy.  In the literature\cite{Wang:2015ndk}, the QCD corrections to the leading power form factors of $\Lambda_b \to \Lambda$ have been calculated, the method can be directly generalized to the $\Lambda_b \to \Lambda_c$ transition. The power suppressed contributions have been shown to be sizable, and a more careful treatment on the power corrections is of great importance. The above mentioned problems will be considered in the future work. 

\section*{Acknowledgement}
We thank Fu-Sheng Yu for very useful discussions and valuable suggestions. 
This work was supported in part by the National Natural Science Foundation of China under
Grant Nos.~12175218,~11975112.
Y.L.S also acknowledges the Natural Science Foundation of Shandong province with Grant No. ZR2020MA093.
J.G is also supported by the National Natural Science Foundation of China under
Grant No.~12147118.

%%%%%%%%%%%%%%%%%%%%%%%%%%%%%%%%%%%%%%%%%%%%%%%%%%%%%%%%%%%%%%%%%%%%%%%%%%%

\end{document}